\documentstyle[aps,epsf]{revtex}

\def\ul#1{\underline{#1}}
\def\kk{\ul{k}}

\def\mat#1{\ul{\ul{#1}}}
\def\intinf{\int\limits_{-\infty}^{\infty}} \def\w{\omega}
\def\ea{e_R}
\def\eb{e_L}
\def\ha{h_R}
\def\hb{h_L}
\def\fb{f_L}

\def\e{\varepsilon}

\def\iwn{i\w_n}

\begin{document}
\twocolumn

\title{Towards a Microscopic Theory for Metallic Heavy-Fermion Point Contacts}
	
\author{F.~B.~Anders$^{1}$ and K.~Gloos$^{2}$}
\address{$^1$ Department of Physics, The Ohio State University, 
 Columbus, Ohio, 43210-1106}
\address{$^2$ Institut f\"ur Festk\"orperphysik,
 Technische Hochschule Darmstadt, D-64289 Darmstadt, Germany}
\date{June 26th, revised July 23rd 1996}

\maketitle

\begin{abstract}
The bias-dependent resistance $R(V)$ of NS-junctions is calculated 
using the Keldysh formalism in all orders of the transfer matrix element. 
We present a compact and simple formula for the Andreev current, that
results from the coupling of electrons and holes on the normal side via 
the anomalous Green's function on the superconducting side. 
Using simple BCS Nambu-Green's functions  the well known
Blonder-Tinkam-Klapwijk theory can be  
recovered. 
Incorporating the energy-dependent quasi-particle lifetime of the heavy 
fermions strongly reduces the Andreev-reflection signal.
\end{abstract}

\section{Introduction}
Point-contact spectroscopy (PCS) has been used to study the superconducting 
(SC) properties of heavy-fermion (HF) compounds, see e.g. Refs. 
~\cite{Nowack87,Dewilde94,Goll95}. The symmetry of the SC order 
parameter in HF compounds is still not known, and it was hoped that PCS is
a useful tool to help clarify this question.  

Usually, the SC anomalies of point contacts between HFSC and normal metals 
have been interpreted in terms of Andreev reflection (AR). But the 
observed spectra deviate significantly from that predicted by the simple 
quasi-classical BTK model \cite{Blonder82}. For example, the doubling 
of the electrical current due to AR at a bias voltage 
$|\text{e}V| = \Delta$ is missing. In recent experiments on UPt$_3$ 
\cite{Goll95} and URu$_2$Si$_2$ \cite{Naidyuk96} (that are good candidates 
for being performed in the ballistic limit as claimed by the authors) 
the size of the SC anomaly amounts to a small fraction of the total signal. 
We present here a microscopic approach using the Keldysh non-equilibrium 
method. It incorporates the strongly energy-dependent lifetime in HF 
compounds, that is also responsible for the large $A$ coefficient of the 
electrical resistivity at temperatures $T\ll T^*$, the lattice Kondo  
temperature.
For vanishing self-energy the BTK result is recovered. 

\section{Theory}
We use a simplified model to describe a NS-junction. Let us assume the 
left side of the junction is described by the Hamiltonian $H_L$, the right 
by $H_R$, and the solution of both is known in terms of the one-particle 
equilibrium Green's function. The two leads are coupled by a tunnel 
Hamiltonian
\cite{SchriefferWil63,Caroli71}
\begin{eqnarray}
\label{equ-hopping}
H_T & = & \sum_{<L,R>} (T_{L,R} c^\dagger_{L\sigma}
c_{R\sigma} + h.c.) \;\; .
\end{eqnarray}
The applied bias voltage translates into a step-like chemical potential at 
the junction with the potential difference $\text{e}V$. Hereby the assumption
of a ballistic point contact enters the theory. For a diffusive or
thermal junction, the  spatial dependence of the chemical potential
has to be determined  
self-consistently. The total current
\begin{equation}\label{equ-ss}
 I(\tau) = \frac{e}{ih} \sum_{L,R}
\mbox{Tr} \left[\tau_3 \left(\mat t \; \mat G_{RL}^{+}(\tau,\tau)
-\mat t^\dagger  
\; \mat
G_{LR}^{+} (\tau,\tau)\right) \right] \;  
\end{equation}
is in general time-dependent. We introduce the Keldysh Nambu $2\times2$ 
occupational Green's function
\begin{equation}
\mat G_{LR}^{+}(\tau,\tau')
=
i \left(
\begin{array}{cc}
\ll c^\dagger_{L\uparrow}(\tau)c_{R\uparrow}(\tau') \gg
&
\ll c^\dagger_{L\uparrow}(\tau)\; c^\dagger_{R\downarrow}(\tau') \gg
\\
\ll c_{L\downarrow}(\tau)\; c_{R\uparrow}(\tau') \gg
&
\ll c_{L\downarrow}(\tau) \; c^\dagger_{R\downarrow}(\tau') \gg
\end{array}
\right)
\end{equation}
with
\begin{equation}
\label{equ-tmat}
\mat t = 
\left(
\begin{array}{cc}
T e^{i\phi_0} & 0 \\
0 & - T e^{-i\phi_0}
\end{array}
\right)
=
\left(
\begin{array}{cc}
t_1 & 0 \\ 0 & -t_1^\star
\end{array}
\right)\;\;\; .
\end{equation}
In this Nambu representation the hopping matrix element picks up a 
time-dependent phase difference $\phi(\tau) = \phi_0 +  \tau\w_0/2$
between the  two leads, with $\w_0 = 2\text{e}V/\hbar$. 

The current can be decomposed in a Fourier series of multiples of the 
fundamental frequency $\w_0$: $I(\tau) = \sum_n I_n \exp(i\tau n\w_0)$
necessary  
for the proper description of a AC Josephson current in SS-junction. In a 
NS-junction, however, all but the $I_0$ component vanish since only $|T|^2$ 
enters in this case. Applying the Keldysh formalism developed for the 
tunneling theory \cite{Caroli71} yields the equations of motion
in frequency space
\begin{eqnarray} 
\mat G^r_{RL}(\w) & = & \mat g^r_{R}(\w) \mat t^\dagger \;
 \mat G^r_{L}(\w)
=
 \mat G^r_{R}(\w) \; \mat t^\dagger \;
 \mat g^r_{L}(\w)
\\
\label{equ-abp}
\mat G^+_{RL}(\w) & = & 
\mat g^+_{R}(\w)\; \mat t \; \mat G^a_{L}(\w)
+
\mat g^r_{R}(\w)\; \mat t \; \mat G^+_{L}(\w)
\end{eqnarray}
which can be solved, omitting the frequency argument  and
 using
\begin{eqnarray}
\label{equ-g-ret}
\mat G_{L}^{r} & = & \mat g_{L}^r + \mat g_{L}^r
\; \mat{\tilde{g}}_{R}^r \;  \mat G_{L}^r \;\;
\Longrightarrow \\ 
 \mat G_{L}^{r} & = &  \frac{1}{1 - \mat g_{L}^r
\; \mat{\tilde{g}}_{R}^r }  \mat g_{L}^r 
\; \mbox{and} \;
 \mat G_{L}^{a} =    \mat g_{L}^a \frac{1}{ 1 - 
 \mat{\tilde{g}}_{R}^a \; \mat g_{L}^a} \;\; ,
\end{eqnarray}
where the two hopping matrix elements have been absorbed into 
the renormalized normal-state propagator $\mat{ \tilde g}_{R}(\w)  \equiv
|T|^2\; \mat g_{R}(\w)$ on the right side (R). The left side (L) becomes SC.
The denominator indicates the re-summation of tunneling processes in
infinite order. 

The occupational-components are more complicated:
\begin{eqnarray}
\mat G_{L}^{\pm} 
& = & 
 \frac{1}{ 1 - \mat g_{L}^r
\; \mat{\tilde{g}}_{R}^r }
\left[
\mat g_{L}^\pm
+ \mat g_{L}^r \mat{\tilde g}_{R}^\pm \mat g_{L}^a
\right]\; \frac{1}{ 1 - 
 \mat{\tilde{g}}_{R}^a \; \mat g_{L}^a }
\; \; .
\label{equ-full-gf}
\end{eqnarray}
Here $g$ denotes the equilibrium advanced (a), retarded (r), and 
occupational ($\pm$) Green's functions in absence of $H_T$, $G$ the fully 
renormalized non-equilibrium Green's functions. Using 
Eqs.~(\ref{equ-abp})-(\ref{equ-full-gf}), the notation $e(\w)$ for an 
electron, $h(\w)$ for a hole, and $f(\w)$ for the anomalous 
Nambu Green's function component, the total DC current $I(V,t=0)$ of 
Eq.(\ref{equ-ss}) decomposes into a quasi-particle current 
\begin{eqnarray}
I_{QP}  &=   &
 \frac{2 e }{h} \sum_{<L,R>} \intinf \; d\w
\frac{T^2}{|Det[ 1 - \mat g_{L}^r \; \mat{\tilde{g}}_{R}^r ]|^2} \nonumber \\
&&  
\left[
| 1 - T^2 \ha^a\hb^a|^2 (\ea^+\eb^- -\ea^-\eb^+) \right.
\nonumber \\
&&  
+
|T^2\ha^a\fb^a |^2 (\ea^+\hb^- -\ea^-\hb^+) \nonumber \\
&& 
\left.
+ 2\Re e[( 1 - T^2 \ha^a\hb^a) T^2\ha^r\fb^r] (\ea^+\fb^- -\ea^-\fb^+) 
 \right]
\label{equ-j-ns-qp}
\end{eqnarray}
and the Andreev current 
\begin{eqnarray}
I_{A} & = &
 \frac{2e}{h}  \sum_{<L,R>} \intinf \; d\w
\frac{T^4 |\fb^r|^2
\left[ (\ea^+\ha^- -\ea^-\ha^+ \right] 
}{|Det[ 1 - \mat g_{L}^r \; \mat{\tilde{g}}_{R}^r ]|^2} 
\label{equ-supercurr}
\end{eqnarray}
with
${|Det[ 1 - \mat g_{L}^r \; \mat{\tilde{g}}_{R}^r ]|^2}  =  |( 1 - T^2 
\ha^r\hb^r) 
( 1 - T^2 \ea^r\eb^r) - T^4\ha^r\ea^r (\fb^r)^2|^2$.

$I_{QP}$ describes three different processes: an electron on the normal 
side R couples to an electron, a hole, or to the anomalous GF. The Andreev 
current couples an electron and a hole on the normal side under the 
influence of a Cooper pair. Its leading order is $T^4$ since it involves 
two quasi-particles crossing the junction. Note that since we resummed all 
orders, the hopping $T$ can be tuned continuously from being much smaller 
than the bandwidth $W$, the tunneling regime, to 
$\gamma = \pi T\, N_F \approx 1$, the metallic PCS regime. 

$\sum_{<L,R>}$ denotes two solvable quasi-1d cases: 
\\
{\em 1. Quantum point contact}.
The orifice involves only a few lattice sites. Translational
invariance is broken and no momentum component is conserved at the
boundary. It can be verified that using BCS propagators and identifying
$Z^2 =((1-\gamma^2)/2\gamma)^2$ the BTK theory is reproduced by
Eqs.~(\ref{equ-j-ns-qp}) and (\ref{equ-supercurr}). where $
\sum_{<L,R>}$ counts the number of channels. For example, the Andreev
current per channel reads 
\begin{eqnarray}
I_A & = &
  \frac{2e}{h} \intinf \;d\w \;T_A(\w) \;\tanh\left[\beta\frac{\w+V}{2} \right]
\\
\left.T_A(\w)\right|_{\w^2<\Delta^2} 
&=& 
\frac{4\gamma^4}{(1+\gamma^4)^2(1-(\w/\Delta)^2) +
4\gamma^4(\w/\Delta)^2}
\\
\left.T_A(\w)\right|_{\w^2>\Delta^2} 
& =& 
\frac{4\gamma^4}{\left((1+\gamma^4)\sqrt{\left(\frac{\w}{\Delta}\right)^2
-1} +2\gamma^2|\w/\Delta|\right)^2}
\end{eqnarray}
Obviously the transmission coefficient $T(\w)$ is continuous at
$\w^2=\Delta^2$ and its value $\frac{2e}{h}$ is independent of the
effective coupling $\gamma$.
\\
{\em 2. Two connected half spaces}.
In this case the momentum parallel to the $\ul{k}_\parallel$-plane is
conserved. $e(\w,\ul{k}_\parallel),f(\w,\ul{k}_\parallel)$ and
$f(\w,\ul{k}_\parallel)$ become $k_z$-summed Green's functions entering
Eq.~(\ref{equ-j-ns-qp}) and (\ref{equ-supercurr}). The total current
is the sum of all independent $\ul{k}_\parallel$ 
contributions. For example, in Eq. \ref{equ-supercurr} an electron is
coupled to a hole of the same $\ul{k}_\parallel$ which translates into
the standard one-particle picture of $\ul{k} \rightarrow -\ul{k}$.
 
\begin{figure}
 \epsfxsize 90mm
 \epsffile{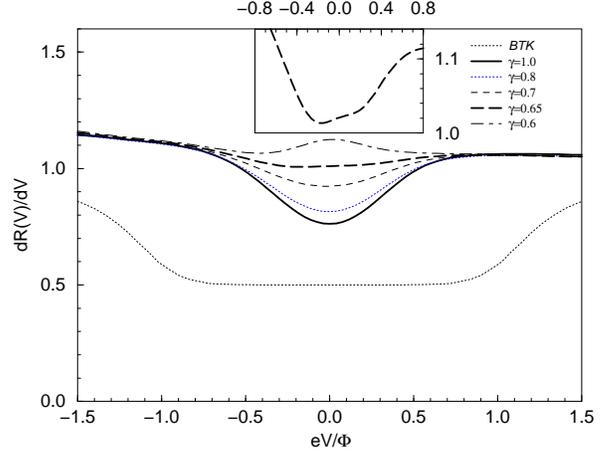}
\caption{Resistance per conductance channel $dV/dI$ vs bias voltage $V$ as 
function of the dimensionless hopping parameter $\gamma$. 
Inset: blow-up of the $\gamma=0.65$ curve.}
\label{fig-1}
\end{figure}

\section{Application}
A microscopic theory of HFSC including quasi-particle lifetime effects is 
missing up to now. We have used instead the normal state Green's function for 
the Anderson lattice
 \begin{equation}
\label{equ-gf-hf}
 \mat g(z) = \frac{1}{N}\sum_{\kk} 
 \left(
 \begin{array}{c}
 z - \e_{\kk} - \frac{V^2}{z-T^\star} - \Sigma_c(z) \;\; , \;\; - \Phi \\ -
 \Phi \;\; ,\;\; z + \e_{-\kk} - \frac{V^2}{z-T^\star} - \Sigma^\ast_c(z) 
 \end{array}
 \right)^{-1}
 \end{equation}
 with an isotropic order parameter $\Phi = 0.1\cdot T^\ast$ as a simple, not 
 self-consistent approximation for the SC HF phase where
$ \Im m \Sigma_{c}(\w-i\delta) = \frac{1}{2\pi N_F}\left( 
 \left(\frac{\pi T}{T^\star} \right)^2 +\left(\frac{\w}{T^\star}\right)^2 
\right)
$. 
This formalism can be easily extended by an anisotropic order parameter 
$\Phi_{\kk}(\iwn)$ using {\em case 2}. Only the conduction
electrons contribute to the charge transport since the
hybridisation is assumed $k$-independent \cite{CoxGre88}.

The results of Eqs.~(\ref{equ-j-ns-qp}) and (\ref{equ-supercurr}) are shown 
in Fig.~\ref{fig-1}. 
Ignoring the life-time effects and the hybridisation $V^2$ in
(\ref{equ-gf-hf}) indeed reproduces the BTK result for $\gamma=1$. But for
the HF case a strongly reduced 'V'-shape SC anomaly  is
found for an ideal metallic point contact ($\gamma =1$). The 'worse'
the contact ($\gamma<1$) the more BTK-like is the shape of an 'ideal'
contact, but its size is strongly reduced at zero bias. Because 
we have chosen a fixed order parameter $\Phi$ the
renormalization due to the strongly reduced quasi-particle spectral
weight leads to the 'shrinking' of the gap. The inset shows a blow-up of
the $\gamma=0.65$ curve. This would come close to some of the spectra
observed for UPt$_3$. The asymmetry reflects
the asymmetry between the normal electrode (treated by a constant 
density of states) and the HFSC. 

\section{Discussion}
According to our above calculations the almost flat spectra with small 
double-minimum structure are accessible only within a very narrow parameter 
range when coupling is reduced to about $\gamma \approx 0.65$. 
This could be one of the reasons why those spectra are very hard to detect 
\cite{Goll95,Naidyuk96}. Note that even with an isotropic order parameter 
spectra different from the BTK predictions can be obtained using a energy
dependent quasi-particle life-time. 

We ought to mention some drawbacks to our approach: 
The energy-dependence of the quasi-particle lifetime requires a 
self-consistent determination of the spatial-dependent potential to 
model contacts in the easily accessible resistance range
($0.3\Omega R < 3 \Omega$). In an Eliashberg-type of calculation
of the SC state the onset of quasi-particle lifetime
effects is expected to be shifted self-consistently towards the
gap-edges.
And we cannot explain the often found enhancement of the ballistic 
zero-bias resistance. 

A recent systematic study on the size of the SC anomalies clearly shows 
a scaling with the orifice radius rather than with the area
\cite{Gloos96-UBe13}. This could indicate that a NNS-junction is
seen in most of the experiments with a NS-boundary inside the HF
material. Such a scenario would be also favoured in the case of an anisotropic
order parameter, since the surface then acts as a pair-braking boundary.
A reminiscence  of SC is found in the
normal-state HF part of the junction, since it is known - or can be
easily derived using Eq.(\ref{equ-g-ret}) -  that the 
non-equilibrium renormalized normal side has a  gap at $\w=0$ in its
spectral function induced by the superconductor it is coupled to, which
is maintained through large distances (in a pure quasi-particle
picture this distance is infinite).

 As long as both the experimental {\em and} the theoretical situation is not 
 solved completely, PCS is unsuitable as 'smoking gun'-technique to 
 determine the symmetry of the HFSC order parameter.

After finishing our work we became aware, that other authors
\cite{MartinRodero96} have also derived the BTK-theory using an
Hamiltonian approach to PCS.
This work has been supported by the Sonderforschungsbereich 252
Darmstadt-Mainz-Frankfurt,  the Deutsche
Forschungsgemeinschaft and in parts by the National Science Foundation
under Grant No.~PHY94-07194. One of us (FBA) likes to thank the ITP,
Santa Barbara for its hospitality.


\end{document}